\documentclass[12pt]{article}
\usepackage{amsfonts, amsmath, amsthm, graphics, bbold}

\def\IM{\hbox{\rm{Im}}\,}
\def\RE{\hbox{\rm{Re}}\,}

\def\ddz{{{\rm d}\over{\rm d} z}}
\def\thi{{\vartheta_\infty}} 
\def\eps{\varepsilon}
\def\phi{\varphi}

\def\arg{{\rm arg}}
\def\iff{\Longleftrightarrow}

\newcommand{\ID}{\mathbb{1}}
\newcommand{\complessi}{\mathbb C}
\newcommand {\interi}{\mathbb Z}
\newcommand{\reali}{I\!\!R}

\theoremstyle{definition}
\newtheorem{df}{Definition}

\newtheorem{rmk}[df]{Remark}

\theoremstyle{theorem}
\newtheorem{lm}[df]{Lemma}
\newtheorem{thm}[df]{Theorem}

\newtheorem{prop}[df]{Proposition}

\begin{document}

\title{\bf Irregular isomonodromic deformations for Garnier systems
and Okamoto's canonical transformations}
\author{Marta Mazzocco\thanks{DPMMS, Cambridge University,
Cambridge CB3 0WB, UK.}} 
\date{}
\maketitle

\begin{abstract}
In this paper we describe the Garnier systems as isomonodromic deformation
equations of a linear system with a simple pole at $0$ and a Poincar\'e rank 
two singularity at infinity. We discuss the 
extension of Okamoto's birational canonical transformations to the 
Garnier systems in more than one variable and to the Schlesinger systems.
\end{abstract}

\section{Introduction}
The $n$--variables {\it Garnier systems}\/ ${\cal G}_n$ were introduced in
\cite{Gar1,Gar2}
as isomonodromic deformations equations of scalar differential equations
of the form
\begin{equation}
{{\rm d}^2y\over{\rm d}x^2}+p_1(x){{\rm d}y\over{\rm d}x}+p_2(x)y=0
\label{scal}\end{equation}
with $n$ apparent singularities $\lambda_1,\dots,\lambda_n$ of indices $(0,2)$ 
and $n+3$ pairwise distinct Fuchsian singularities 
$t_1,\dots,t_{n+3}$ (in particular $t_{n+1}=0$, $t_{n+2}=1$, 
$t_{n+3}=\infty$) with indices $(0,\theta_1)$,\dots,$(0,\theta_{n+2})$,
$\left(-{\sum_{k=1}^{n+2}\theta_k +\theta_\infty\over2}, 
-{\sum_{k=1}^{n+2}\theta_k -\theta_\infty\over2}+1\right)$ respectively.
The constraints on the indices are enough to determine $p_1(x)$, $p_2(x)$
in terms of  $\lambda_1,\dots,\lambda_n$ and $n$ other quantities 
$\mu_1,\dots,\mu_n$ (see \cite{IKSY})
\begin{equation}
\begin{array}{l}
p_1(x)=\sum_{k=1}^{n+2}{1-\theta_k\over x-t_k}-
\sum_{i=1}^{n}{1\over x-\lambda_i},\\
p_2(x)={\kappa\over x(x-1)}
-\sum_{i=1}^{n}{t_i(t_i-1)K_i\over x(x-1)(x-t_i)}+
\sum_{i=1}^{n}{\lambda_i(\lambda_i-1)\mu_i\over x(x-1)(x-\lambda_i)},\\
\end{array}
\label{scal1}\end{equation}
where $\kappa={1\over4}\left\{ \left(
\sum_{k=1}^{n+2}\theta_k-1\right)^2-(\thi+1)^2\right\}$ and 
\begin{equation}
K_i=-{\Lambda(t_i)\over T'(t_i)}\left[\sum_{k=1}^{n}
{T(\lambda_k)\over(\lambda_k-t_i)\Lambda'(\lambda_k)}
\left\{\mu_k^2-
\sum_{m=1}^{n+2}{\theta_m-\delta_{im}\over\lambda_k-t_m}\mu_k
+{\kappa\over\lambda_k(\lambda_k-1)}\right\}\right]
\label{ham}\end{equation}
with
$\Lambda(u):=\Pi_{k=1}^{n}(u-\lambda_k)$ and
$T(u):=\Pi_{k=1}^{n+2}(u-t_k)$.

The isomonodromic deformations equations of the above equation are 
described by the following completely integrable Hamiltonian system 
\cite{Gar1,Gar2,Ok1,IKSY}, the {\it Garnier systems}\/ ${\cal G}_n$:
\begin{equation}
\left\{\begin{array}{l}
{\partial\lambda_j\over\partial t_i}={\partial
K_i\over\partial\mu_j}
\qquad i,j=1,\dots,n,\\
{\partial\mu_j\over\partial t_i}=-{\partial K_i\over\partial\lambda_j}
\qquad i,j=1,\dots,n.\\
\end{array}
\right.
\label{ga1}\end{equation}

In this paper we show that the Garnier systems ${\cal G}_n$ can be 
interpreted as isomonodromic deformations equations of the
following $(n+2)\times(n+2)$ linear system with a simple pole at $0$ and a 
Poincar\'e rank two singularity at $\infty$
\begin{equation}
{{\rm d}\over{\rm d}z}Y
=\left(\left(\begin{array}{ccc}t_1& & \\ & \dots&\\ & & t_{n+2}
\end{array}\right)+{V-\ID\over z}\right) Y,
\label{irreg}\end{equation}
where $V$ is a $(n+2)\times(n+2)$ diagonalizable matrix with $n$
null eigenvalues 
\begin{equation}
G^{-1}VG = \rho:=\left(\begin{array}{ccccc}
\rho_1& & & &\\
&\rho_2& & &\\ & & 0 & &\\ & & & \dots & \\ & & & & 0\\
\end{array}\right),
\label{cV}\end{equation}
with $\rho_1={1\over 2}(-\sum\theta_k+\thi)$,
$\rho_2={1\over 2}(-\sum\theta_k-\thi)$. We show that
if $\lambda_1,\dots,\lambda_n$, $\mu_1,\dots,\mu_n$ are solutions of the 
Garnier system (\ref{ga1}), then they uniquely determine the matrix 
$V(t_1,\dots,t_n)$ up to diagonal conjugation in such a way that the 
monodromy data of the system 
$$
{{\rm d}\over{\rm d}z}Y
=\left(\left(\begin{array}{ccc}t_1& & \\ & \dots&\\ & & t_{n+2}
\end{array}\right)+{V(t_1,\dots,t_n)-\ID\over z}\right) Y,
$$ 
are constant. In particular $V$ is given by the 
following  formulae
\begin{equation}
V_{ii} = -\theta_i, \quad i=1,\dots,n+2,
\label{cV1}\end{equation}
\begin{equation}
V_{ij} = -\theta_j+ M_j W_j-M_j W_i,\quad i,j=1,\dots,n+2,
\label{off}\end{equation}
where $M_i, W_i$ are given in terms of the variables 
$(\lambda_1,\dots,\lambda_n,\mu_1,\dots,\mu_n)$
(here we use the same notations of \cite{IKSY}) by:
\begin{equation}
\begin{array}{l}
M_i:=-{\Lambda(t_i)\over T'(t_i)}, \quad i=1,\dots,n+2,\\
W_j:=\sum_{k=1}^n {T(\lambda_k)\over(\lambda_k-t_j)\Lambda'(\lambda_k)} 
\left(\mu_k-{\sum_{k=1}^{n+2}\theta_k+\theta_\infty
\over2\lambda_k}\right), \quad j=1,\dots,n,\\
W_{n+1}={1\over M_{n+1}}\left(\sum_{j=1}^n (t_j-1)M_j W_j+
{\sum_{k=1}^{n+2}\theta_k+\theta_\infty\over2}\right),\\
W_{n+2}=-{1\over M_{n+2}}\sum_{j=1}^n t_j M_j W_j.\\
\end{array}
\label{forj}
\end{equation}
In the case of $n=1$, corresponding to the Painlev\'e sixth
equation, such result was already known \cite{Hr,M3}. Nevertheless 
the generalization to the Garnier systems ${\cal G}_n$ with $n>1$ variables
is non-trivial. 

Our result seems to be of fundamental importance if one wants to
investigate the possibility to extend Okamoto's birational canonical 
transformations defined for the Painlev\'e sixth equation to the case of 
Garnier systems. In \cite{Ok2} it is shown that the group of all birational 
canonical transformations of the Painlev\'e sixth equation is isomorphic to
the affine extension of $W(F_4)$. The generators are given by the 
generators $w_1,w_2,w_3,w_4$ of $W(D_4)$, the parallel transformations 
$l_1,l_2,l_3$, and the symmetries $x_1,x_2, x_3$ (we shall remind 
details of these in Section 3). The symmetries can be generalized to
the case of Garnier systems \cite{IKSY}, and indeed to the Schlesinger systems
in any dimension \cite{DM1}. Recently Tsuda \cite{Ts} defined 
the analogues of $w_1,w_3,w_4,l_1,l_2,l_3$ for the Garnier system and together
with Sakai conjectured that the analogue of $w_2$ does not exist \cite{ST}. 
Here we give further strong evidence that Sakai-Tsuda conjecture is valid.
In fact, in the case of the Painlev\'e sixth equation, $n=1$, we generate 
a birational canonical transformation $\tilde w$ equivalent to $w_2$,
$\tilde w= x_2 w_2 x_2$, as a simple scalar gauge transformation of the 
irregular system (\ref{irreg})
$$
\tilde Y(z)=z^{-\rho} Y(z),\qquad V\to V-\rho\ID,
$$
where $\rho$ is one of the eigenvalues of $V$. In this way the condition
that one of the eigenvalues of $V$ is zero is preserved. The idea is that 
for $n>1$, $V$ has $n>1$ null eigenvalues, and this condition is not 
preserved by the above gauge transformation. Conversely it is possible
to extend $\tilde w$ to a special case of the Schlesinger systems.

A second application of our result is in the theory of Frobenius manifolds
\cite{Dub7}. 
A $\tilde n$-dimensional Frobenius manifolds is locally parameterized 
by the monodromy data of a $\tilde n\times\tilde n$ linear system 
similar to (\ref{irreg}):
\begin{equation}
{{\rm d}\over{\rm d}z}\tilde Y
=\left(\left(\begin{array}{ccc}t_1& & \\ & \dots&\\ & & t_{\tilde n}
\end{array}\right)+{W\over z}\right)\tilde Y,
\label{irregf}\end{equation}
where $W$ is a $\tilde n\times\tilde n$ diagonalizable antisymmetric matrix. 
The eigenvalues $\tilde\rho_1,\dots,\tilde\rho_{\tilde n}$ of $W$ are related 
to the degrees $d_1,\dots,d_{\tilde n}$ of the Frobenius manifold by
$d_i=1-\tilde\rho_i-{d\over2}$. In the special case 
$$
\begin{array}{l}
d_1=1,\quad d_{\tilde n}=1-d,\\
d_i=1-{d\over2},\quad i\neq1,\tilde n,\\
\end{array}
$$
the system (\ref{irregf}) is equivalent to the system (\ref{irreg}) with
$n=\tilde n-2$, $\theta_1=\theta_2=\dots=\theta_{n+2}=0$ and 
$\theta_\infty=d$.\footnote{Such equivalence is realized by $\tilde Y=z Y$. 
The fact that $V$ becomes antisymmetric follows immediately from the formula
(\ref{off}) after a sutable diagonal conjugation.} 
In other words we show that the Garnier system (\ref{ga1}) describes 
the analytic structure of Frobenius manifolds of degrees
$1, 1-d, 1-{d\over2},\dots,1-{d\over2}$ for any $d$.

This paper is organized as follows: in Section 2 we show that the 
Garnier systems ${\cal G}_n$ can be 
interpreted as isomonodromic deformations equations of the
$(n+2)\times(n+2)$ linear system (\ref{irreg}). In Section 3  we give a 
brief resume of Okamoto's results on birational canonical 
transformations for the Painlev\'e sixth equation.
In Section 4 we discuss the isomonodromic meaning of Okamoto's birational 
canonical transformation $w_2$ and its extension to the Schlesinger systems.

\vskip 0.1 cm
\noindent{\bf Acknowledgments} The author is grateful to H. Sakai and T. Tsuda
for helpful discussions and to Prof. Okamoto for kindly inviting her to Tokyo
graduate school of Mathematics where this paper was started.

\section{Isomonodromic interpretation of the Garnier systems}

In this section we prove that the Garnier systems are isomonodromic 
deformations equations of the system (\ref{irreg}). These are described in
the following theorem proved in \cite{MJU,Dub1}:

\begin{thm}
Let $V({\bf t})$, ${\bf t}=(t_1,\dots,t_{n})$, be a diagonalizable matrix 
function with eigenvalues $\rho_1,\rho_2,0,\dots,0$. Suppose that for 
every $i=1,\dots,n$ the matrix function $V({\bf t})$ satisfies
\begin{equation}
{\partial\over\partial t_i}V=[V_i,V],
\label{15}
\end{equation}
where
$$
V_i={\rm ad}_{E_i}{\rm ad}_{\bf T}^{-1}(V),\quad
{\bf T}=\left(\begin{array}{ccc}t_1& & \\ & \dots&\\ & & t_{n+2}
\end{array}\right),
$$
and $E_{i_{kl}}=\delta_{ik}\delta_{il}$. Then the monodromy 
data\footnote{See appendix for a description of the monodromy data of the 
system (\ref{irreg}).} of the system  (\ref{irreg})
$$
{{\rm d}\over{\rm d}z}Y
=\left({\bf T}+{V({\bf{t}})-\ID\over z}\right) Y,
$$
are constant.
\end{thm}

Our main result is the following:

\begin{thm}
Consider the $(n+2)\times(n+2)$ diagonalizable matrix $V({\bf t})$, 
${\bf t}=(t_1,\dots,t_{n})$, satisfying the conditions (\ref{cV})
and (\ref{cV1}). Then its off diagonal entries can be expressed in 
terms of the variables $\lambda_1,\dots,\lambda_n,\mu_1,\dots,\mu_n$ 
by (\ref{off}), where $M_i$ and $W_i$ are given in (\ref{forj}). Moreover
$V({\bf t})$ satisfies (\ref{15}) if and only if 
$\lambda_1,\dots,\lambda_n,\mu_1,
\dots,\mu_n$ satisfy the Garnier system ${\cal G}_n$ given in (\ref{ga1}).
\label{main}\end{thm}

\noindent{\bf Proof.}
The idea of the proof is to show that the systems of the form (\ref{irreg})
satisfying the conditions (\ref{cV}) and (\ref{cV1}),
 are equivalent to $2\times 2$ Fuchsian systems of the form
\begin{equation}
{{\rm d}\over{\rm d}x} \Phi=\left(
\sum_{k=1}^{n+2}{{A}_k\over x-t_k} \right)\Phi,
\label{N1in}
\end{equation}
where the residue matrices ${A}_k$ satisfy the following 
conditions:
\begin{equation}
{\rm eigen}\left({A}_k\right)=(0,{\theta_k})
\quad\hbox{and}\quad
-\sum_{k=1}^{n+2}{A}_k={A}_\infty:=\left(
\begin{array}{cc}
\rho_1 & 0\\ 0 & \rho_2\\
\end{array}\right),
\label{N1.3}\end{equation}
with $\rho_1={1\over 2}(-\sum\theta_k+\thi)$, 
$\rho_2={1\over 2}(-\sum\theta_k-\thi)$.
We then shall use the fact the Garnier systems are isomonodromic deformations
equations for such $2\times 2$ Fuchsian systems.
\vskip 0.3 cm

More precisely, we introduce the following

\begin{df} Two $(n+2)\times(n+2)$ systems of the form (\ref{irreg}) are 
{\it equivalent up to diagonal conjugation}\/ if they have the 
same matrix ${\bf T}$ and the same matrix $V$ up to
$V\to D_{n+2} V D_{n+2}^{-1}$, $D_{n+2}$ being any $(n+2)\times(n+2)$
diagonal matrix. 
Analogously, two $2\times2$ Fuchsian systems of the form (\ref{N1in}) 
are {\it equivalent up to diagonal conjugation}\/ if they have the 
same matrices ${A}_k$, $k=1,2,3$, up to
${A}_k\to D_2^{-1} {A}_k D_2$, $D_2$ being any diagonal matrix.
\label{dfequi}\end{df}

\begin{lm} 
There is a one to one correspondence between classes of equivalence of 
$(n+2)\times(n+2)$ systems of the form (\ref{irreg})
$$
{{\rm d}\over{\rm d}z}Y
=\left({\bf T}+{V-\ID\over z}\right) Y,
$$
where $V$ is a $(n+2)\times(n+2)$ diagonalizable matrix satisfying the 
conditions (\ref{cV}) and (\ref{cV1})
and classes of equivalence of $2\times2$ systems of the form (\ref{N1in})
$$
{{\rm d}\over{\rm d} x} \Phi
=\sum_{k=1}^{n+2} {{A}_k\over x-t_k}\Phi 
$$
where ${A}_k$ satisfy (\ref{N1.3}).
\label{lmequiv}\end{lm}

\noindent{\bf Proof.} Consider the following $(n+2)\times(n+2)$ Fuchsian system
\begin{equation}
{{\rm d}\chi\over{\rm d}x}=(U-x\ID)^{-1}V\chi=\left(
\sum_{k=1}^{n+2}{{\cal B}_k\over x-t_k} \right)\chi,
\label{schi}
\end{equation}
where ${\cal B}_k=-E_k V$, i.e. $\left({\cal B}_k\right)_{ij}=
-\delta_{ik} V_{ij}$. Since $V$ is a diagonalizable $(n+2)\times(n+2)$ matrix
with $n$ null eigenvalues, the gauge transformation $\chi=G\tilde\chi$,
where $G^{-1}VG=\rho:={\rm diagonal}(\rho_1,\rho_2,0,\dots,0)$ maps 
(\ref{schi}) to
\begin{equation}
{{\rm d}\tilde\chi\over{\rm d}x}=\left(
\sum_{k=1}^{n+2}{\tilde{\cal B}_k\over x-t_k} \right)\tilde\chi,
\label{schit}
\end{equation}
where $\tilde{\cal B}_k=-G^{-1} E_k G\,\rho$,
i.e. all matrices $\tilde{\cal B}_k$ have all last $n$ columns equal to zero.
The system (\ref{schit}) therefore reduces to a $2\times2$ Fuchsian system
of the form (\ref{N1in})
for the first two rows of $\tilde\chi$, 
$\Phi=\left(\begin{array}{c}\tilde\chi_1\\ \tilde\chi_2\\
\end{array}\right)$, with $2\times 2$ residue matrices 
\begin{equation}
{A}_k=\left(\begin{array}{cc}
(\tilde{\cal B}_k)_{11}&(\tilde{\cal B}_k)_{12}\\
(\tilde{\cal B}_k)_{21}&(\tilde{\cal B}_k)_{22}\\
\end{array}\right).
\label{matA}\end{equation}
By construction, the residue matrices ${A}_k$ satisfy (\ref{N1.3}), 
where $\rho_1={1\over 2}(-\sum\theta_k+\thi)$, 
$\rho_2={1\over 2}(-\sum\theta_k-\thi)$ are the two non-null eigenvalues of 
$V$. 

It is very easy to verify that the off--diagonal elements of $V$ satisfy
the following relation
$$
V_{kl} V_{lk} = {\rm Tr}({\mathcal B}_k{\mathcal B}_l) =
{\rm Tr}(\tilde{\mathcal B}_k\tilde{\mathcal B}_l) =
{\rm Tr}({A}_k{A}_l),
$$
therefore given ${A}_1,\dots,{A}_{n+2}$, we determine
$V$ up to diagonal conjugation. 
Vice versa given $V$, it uniquely determines ${\cal B}_k=-E_k V$ and $G$
up to $G\to G S$, where $S$ is a $(n+2)\times(n+2)$ invertible matrix such that
$S_{12}=S_{21}=0$ and $(S^{-1})_{11}={1\over S_{11}}$, $(S^{-1})_{22}=
{1\over S_{22}}$, $(S^{-1})_{12}=0$, $(S^{-1})_{21}=0$. This determines 
$\tilde{\cal B}_k=G^{-1}{\cal B}_kG$
up to $\tilde{\cal B}_k\to S^{-1}\tilde{\cal B}_k S$, therefore  
$V$ uniquely determines ${A}_k$
up to diagonal conjugation.{\hfill $\bigtriangleup$}
\vskip 0.3 cm

The isomonodromic deformations equations of the Fuchsian system (\ref{N1in})
are described in the following theorem proved in \cite{Sch}.

\begin{thm}
Let $A_1({\bf t}),\dots,A_{n+2}({\bf t})$, ${\bf t}=(t_1,\dots,t_{n})$, 
be diagonalizable matrices satisfying (\ref{N1.3}). Suppose that for 
every $i=1,\dots,n$ the matrix functions $A_1({\bf t}),\dots,A_{n+2}({\bf t})$
satisfy
\begin{equation}
\left\{\begin{array}{l}
{\partial\over\partial t_j} {A}_i= 
{[ {A}_i, {A}_j]\over t_i-t_j},\qquad i\neq j,\\
{\partial\over\partial t_i} {A}_i= 
-\sum_{j\neq i}{[ {A}_i, {A}_j]\over t_i-t_j}.\\
\end{array}\right. 
\label{schl}
\end{equation}
Then the monodromy data of 
the system  
$$
{{\rm d}\over{\rm d} x} \Phi
=\sum_{k=1}^{n+2} {{A}_k\over x-t_k}\Phi 
$$
are constant.
\end{thm}

The relation between Schlesinger equations and Garnier systems is given in 
the following

\begin{lm}\cite{IKSY}
Let $\lambda_1,\dots,\lambda_n$ be the roots of the equation
$$
\sum_{k=1}^{n+2} {{A}_{k_{12}}\over x-t_k}=0
$$ 
and define $\mu_1,\dots,\mu_n$ as
$$
\mu_i:=\sum_{k=1}^{n+2} {{A}_{k_{11}}\over \lambda_i-t_k},\quad i=1,\dots,n.
$$ 
Then $A_1,\dots,A_{n+2}$ are uniquely determined up to diagonal conjugation
by $\lambda_1,\dots,\lambda_n,\mu_1,\dots,\mu_n$ 
\begin{equation}
\begin{array}{l}
A_{k_{11}}= M_k(W_k-W)\\
A_{k_{12}}=-M_i \\
A_{k_{21}}= -(W_k-W)[ M_k(W-W_k)+\theta_k]\\
A_{k_{22}}= \theta_k-M_k(W_k-W),\\
\end{array}
\label{fA}\end{equation}
where $M_k$ and $W_k$ are given in terms of the 
variables $\lambda_1,\dots,\lambda_n,\mu_1,\dots,\mu_n$ by (\ref{forj}) and
$\theta_\infty W=\sum_{j=1}^{n+2}W_j(M_j W_j-\theta_j)$. Moreover 
$A_1,\dots,A_{n+2}$ satisfy the Schlesinger equations if and only if 
$\lambda_1,\dots,\lambda_n,\mu_1,\dots,\mu_n$ satisfy the Garnier 
system ${\cal G}_n$ given in (\ref{ga1}).
\end{lm}

In particular, using the formulae (\ref{fA}) for the matrices 
${A}_1,\dots,{A}_{n+2}$ and we obtain that the off-diagonal
elements of $V$ are given by the formulae (\ref{off}) up to diagonal 
conjugation.

In order to complete the proof of Theorem \ref{main} it is enough to prove the
following

\begin{lm}
The matrix $V({\bf t})$, ${\bf t}=(t_1,\dots,t_{n})$ satisfies (\ref{15})
if and only if the matrices $A_1,\dots,A_{n+2}$ defined by (\ref{matA})
satisfy the Schlesinger equations. \label{lmequiv1} \end{lm}

\noindent{\bf Proof.}
The matrix $V({\bf t})$, ${\bf t}=(t_1,\dots,t_{n})$ satisfies (\ref{15})
if and only if the diagonalizing matrix $G$ such that $G^{-1}VG=\rho$, 
satisfies
\begin{equation}
\left[ G^{-1}{\partial G\over\partial t_i},\rho\right]=
\left[G^{-1} V_i G, \rho\right],\qquad \forall i.
\label{a}\end{equation}
This equation is valid if and only if 
$ G^{-1}{\partial G\over\partial t_i} =G^{-1}V_iG +T_i$,
where $T_i$ is a $(n+2)\times(n+2)$ matrix such that $T_{i_{12}}=T_{i_{21}}=0$.
Let us define ${\cal A}_k:=\Pi\tilde{\cal B}_k$, where 
$\Pi={\rm diagonal}(1,1,0,\dots,0)$. It is clear that $A_1,\dots,A_{n+2}$ 
satisfy the Schlesinger equations if and only if ${\cal A}_1,\dots,
{\cal A}_{n+2}$ do. This happens if and only if
$$
\Pi\left[G^{-1}{\partial G\over\partial t_i} ,G^{-1}E_kG\right]\rho=
 \Pi\left[G^{-1} V_i G,G^{-1} E_kG\right]\rho,\qquad \forall\,k\neq i,
$$
and
\begin{equation}
\Pi\left[G^{-1}{\partial G\over\partial t_i} ,G^{-1}E_iG\right]\rho=
-\sum_{k\neq i} \Pi\left[G^{-1} V_i,G^{-1} E_kG\right]\rho.
\label{b}\end{equation}
These equations are valid if and only if $ G^{-1}{\partial G\over\partial t_i} 
=G^{-1}V_iG +R_i$, where $R_i$ is a $(n+2)\times(n+2)$ matrix such that 
$\Pi[R_i,G^{-1} E_kG]\rho=0$ for every $k\neq i$. Since $G$ is determined by
$V$ up to  $G\to G S$, where $S$ is a $(n+2)\times(n+2)$ matrix such that
$S_{12}=S_{21}=0$, and $(S^{-1})_{11}={1\over S_{11}}$, $(S^{-1})_{22}=
{1\over S_{22}}$, $(S^{-1})_{12}=0$, $(S^{-1})_{21}=0$, we can choose 
$G$ in such a way that 
$\Pi[R_i,G^{-1} E_kG]\rho$ is zero if and only if $R_i=T_i$. This shows that 
(\ref{a}) is valid if and only if (\ref{b}) is.
{\hfill $\bigtriangleup$}

\section{Birational canonical transformations for the Painlev\'e sixth 
equation}

For $n=1$, the Garnier system becomes and ODE in the variable $t_1=t$,
the Painlev\'e sixth equation
$$
\begin{array}{ll}
{{\rm d}^2\lambda\over{\rm d}t^2}=&{1\over2}\left({1\over \lambda}+
{1\over \lambda-1}+{1\over \lambda-t}\right)
\left({{\rm d}^2\lambda\over{\rm d}t^2}\right)^2 -
\left({1\over t}+{1\over t-1}+
{1\over \lambda-t}\right){{\rm d}^2\lambda\over{\rm d}t^2}+\\
&+{\lambda(\lambda-1)(\lambda-t)\over t^2(t-1)^2}
\left[\alpha+\beta {t\over \lambda^2}+
\gamma{t-1\over(\lambda-1)^2}+
\delta {t(t-1)\over(\lambda-t)^2}\right],\\
\end{array}
$$
where
\begin{equation}
\alpha={(\theta_\infty-1)^2\over2},\quad\beta=-{\theta_2^2\over2},
\quad\gamma={\theta_3^2\over2},\quad\delta={1-\theta_1^2\over2}.
\label{pa}\end{equation}
In Okamoto's papers the parameters 
\begin{equation}
\kappa_0=\theta_2,\quad \kappa_1=\theta_3,\quad  
\kappa_\infty=\theta_\infty-1,\quad 
\theta=\theta_1,
\label{relo}\end{equation}
are used. To fix notations, and for convenience of the reader, let us 
remind the results of \cite{Ok2}. In the canonical variables $(\lambda,\mu)$, 
the Hamiltonian function $K$ reads:
\begin{equation}
\begin{array}{ll}
K= & {1\over  t(t-1)} \big\{\lambda(\lambda-1)(\lambda-t) \mu^2
+b_3 b_4 (\lambda-t)-\\
&-\left[(b_1+b_2)(\lambda-1)(\lambda-t)+(b_1-b_2)\lambda(\lambda-t)+
(b_3+b_4)\lambda(\lambda-1)\right] \mu\big\}\\ 
\end{array}\label{2.1}
\end{equation}
where the parameters $(b_1,b_2,b_3,b_4)$ are given by
\begin{equation}
b_1={\kappa_0+\kappa_1\over 2},\quad
b_2={\kappa_0-\kappa_1\over 2},\quad   
b_3={\theta-1+\kappa_\infty\over 2},\quad
b_4={\theta-1-\kappa_\infty\over 2}.\label{2.2}
\end{equation}
Okamoto's birational canonical transformations are given in terms of 
transformations of a certain auxiliary Hamiltonian function
$$
h(t)=t(t-1)K(t)+(b_1 b_3+b_1 b_4 + b_3 b_4) t -
{b_1 b_2 + b_1 b_3+b_1 b_4 +b_2 b_3+b_2 b_4 + b_3 b_4\over 2},\label{2.3}
$$
that satisfies the following nonlinear ordinary
differential equation
\begin{equation}
{{\rm d} h\over{\rm d} t}
\left[t(1-t){{\rm d}^2 h\over{\rm d} t^2}\right]^2+
\left\{{{\rm d} h\over{\rm d} t}
\left[2 h-(2t-1){{\rm d} h\over{\rm d} t}\right]+ b_1 b_2 b_3 b_4\right\}^2
=\prod_{i=1}^4 \left({{\rm d} h\over{\rm d} t}+b_i^2 \right).
\label{2.4}
\end{equation}
A {\it singular solution}\/ $h$ to (\ref{2.4}) is a solution linear in $t$:
$$
h(t)= a t +b,
$$
for some constants $a$ and $b$. For the case $\theta_{1,2,3}=0$, these
singular solutions coincide with $\lambda=t_i$ and $\lambda=\infty$. Okamoto 
(see \cite{Ok2})
proves the following:
 
\begin{lm} There is a one-to-one correspondence between non
singular solutions $h$ to (\ref{2.4}) and solutions $(\mu,\lambda)$ to the
Hamiltonian system with Hamiltonian function $K$ given by (\ref{2.1}),
or equivalently solutions $\lambda$ of the Painlev\'e VI equation with 
parameters $\alpha,\beta,\gamma,\delta$ given by (\ref{pa}).
\end{lm}
 
Such a correspondence is realized by the formulae at page 354 of 
\cite{Ok2}. The
canonical transformations of the general PVI are all listed by Okamoto and 
consist of three families:
\begin{enumerate}
\item Generators $w_1,w_2,w_3,w_4$ of $W(D_4)$, which leave the equation 
(\ref{2.4}) invariant
$$
\begin{array}{ll}
w_1(b_1, b_2, b_3, b_4)&=(b_2, b_1, b_3, b_4),\\
w_2(b_1, b_2, b_3, b_4)&=(b_1, b_3, b_2, b_4),\\
w_3(b_1, b_2, b_3, b_4)&=(b_1, b_2, b_4, b_3),\\ 
w_4(b_1, b_2, b_3, b_4)&=(b_1, b_2, -b_3, -b_4).\\
\end{array}
$$
The action of the transformations $w_1,\dots,w_4$ on $\mu,\lambda$ is given 
by formula (2.10) in \cite{Ok2}.
\item Parallel transformations $l_i$, which change the auxiliary 
Hamiltonian $h$.
They act on the parameters as follows:
$$
l_i(b_j)=b_j\quad\hbox{for }\, j\neq i,\qquad l_i(b_i)=b_i+1
$$
and on the auxiliary Hamiltonian $h$ as:
$$
l_i(h(b_1, b_2, b_3, b_4))=h(l_i(b_1, b_2, b_3, b_4)).
$$
\item The symmetries $x_i$, which change also the variable $t$:
$$
\begin{array}{ll}
x_1(\mu,\lambda,t,b_1, \dots, b_4)=
&(-\mu,1-\lambda,1-t,b_1,-b_2,b_3,b_4),\\
x_2(\mu,\lambda,t,b_1, \dots, b_4)=
&\big((b_1+b_3) \lambda-\lambda^2 \mu,{1\over \lambda},{1\over t},
{b_1-b_2+b_3-b_4\over2},{b_2-b_1+b_3-b_4\over2},\\
&{b_1+b_2+b_3+b_4\over2}{-b_1-b_2+b_3+b_4\over2}\big),\\
x_3(\mu,\lambda,t,b_1,\dots, b_4)
=&\big(-(t-1)\mu,{\lambda-t\over 1-t},{1\over t-1},
{b_1-b_2+b_3+b_4+1\over2},{b_2-b_1+b_3+b_4+1\over2},\\
&{b_1+b_2+b_3-b_4-1\over2},{b_1+b_2-b_3+b_4-1\over2}
\big).\\ \end{array}
$$
\end{enumerate}
Okamoto proves that all these transformations are realized as birational 
canonical
transformations of $(\mu,\lambda)$, provided that the correspondent auxiliary
Hamiltonian is non singular.

\section{Okamoto's $w_2$ transformation} 

As we mentioned in the introduction, the symmetries $x_1,x_2,x_3$ can all 
be obtained as conformal transformations on the variable $x$ of the linear 
differential equation (\ref{scal}). This fact can be extended to Garnier 
systems as explained in \cite{IKSY} and to Schlesinger systems in any 
dimensions \cite{DM1}. All transformations $w_i$ apart from 
$w_2$ can be obtained as constant gauge transformations on the 
Fuchsian system (\ref{N1in}), that can be generalized to Garnier systems 
\cite{Ts}. The affine transformations can be obtained 
as Schlesinger transformations on the Fuchsian system and can also be 
generalized to Garnier systems \cite{Ts}. The isomonodromic meaning of $w_2$
was until now unknown, we explain it here below using the irregular system 
(\ref{irreg}). We actually explain the meaning of an equivalent 
transformation $\tilde w=x_2 w_2 x_2$ such that 
$$
\tilde w(\mu,\lambda,t,b_1,b_2,b_3,b_4) = 
(\tilde\mu,\tilde\lambda,t,-b_4,b_2,b_3,-b_1),
$$
where $(\tilde\mu,\tilde\lambda)$ can be obtained by formula (2.10) in 
\cite{Ok2}. Our transformation $\tilde w$ is obtained by the following 
simple gauge transformation of the irregular system (\ref{irreg})
$$
\tilde Y(z)=z^\gamma Y(z),\qquad 
\gamma={\theta_1+\theta_2+\theta_3-\theta_\infty\over2}=b_1+b_4= 
-\rho_1. 
$$
In fact this gauge transformation maps the system (\ref{irreg}) to
$$
{{\rm d}\over{\rm d}z}\tilde Y
=\left(\left(\begin{array}{ccc}t_1& & \\ & \dots&\\ & & t_{n+2}
\end{array}\right)+{\tilde V-\ID\over z}\right) \tilde Y,
$$
where $\tilde V=V-\rho\ID$. This gauge transformation is therefore compatible 
with the condition that $V$ has one null eigenvalue. In fact 
$$
\begin{array}{l}
\theta_1\to\tilde\theta_1=\theta_1+\rho_1,\\
\theta_2\to\tilde\theta_2=\theta_2+\rho_1,\\
\theta_3\to\tilde\theta_3=\theta_3+\rho_1,\\
\theta_\infty\to\tilde\theta_\infty=\theta_\infty-\rho_1,\\
\rho_1\to 0,\\
\rho_2\to\tilde\rho_2=\rho_2-\rho_1,\\
0\to\tilde\rho_1=-\rho_1.
\end{array}
$$
Using (\ref{relo}) and (\ref{2.2}), we immediately obtain that such gauge 
transformation maps $(b_1,b_2,b_3,b_4)$ to $(-b_4,b_2,b_3,-b_1)$. The
transformation law for $(\tilde\mu,\tilde\lambda)$ follows immediately from 
the fact that $V$ satisfies (\ref{15}) if and only if $\tilde V$ does. Then
it is possible to apply the same procedure as in Section 2 to $\tilde V$, i.e.
express $\tilde V$ in terms of $\tilde\lambda$, $\tilde\mu$, solutions of the
Garnier system (Painlev\'e sixth equation) in one variable $t$ with parameters
$\tilde\theta_1,\tilde\theta_2,\tilde\theta_3,\tilde\theta_\infty$. Then
$(\tilde\mu,\tilde\lambda)$ must be related to $(\mu,\lambda)$ by formula 
(2.10) in \cite{Ok2}.
It is clear that in the case of Garnier systems with $n>1$ variables,
the condition that $V$ has $n>1$ null eigenvalues is not 
preserved by the above gauge transformation. Therefore we expect that
the analogous of the transformation $\tilde w$ or equivalently of $w_2$ does 
not exist for Garnier systems with $n>1$ variables \cite{ST}. Conversely, we
expect such a transformation to extend to the special cases of Schlesinger 
systems (\ref{schl}) as explained here below.

\begin{thm}
Let $m$ be any integer $0<m<n+2$, and let $\rho_1,\dots,\rho_{m}$ pairwise 
distinct non-zero complex numbers. There is a one-to-one correspondence 
between equivalence classes of $(n+2)\times(n+2)$ systems of the form 
(\ref{irreg}) where $V$ is a $(n+2)\times(n+2)$ diagonalizable matrix 
satisfying the conditions (\ref{cV}) and 
\begin{equation}
G^{-1}VG = \rho:=\left(\begin{array}{ccccc}
\rho_1& & & &\\
&\dots & & &\\ & & \rho_m & &\\ & &  & 0  & \\ & & & & 0\\
\end{array}\right),
\label{cV2}\end{equation}
and classes of equivalence of $m\times m$ systems of the form (\ref{N1in})
$$
{{\rm d}\over{\rm d} x} \Phi
=\sum_{k=1}^{n+2} {{A}_k\over x-t_k}\Phi 
$$
where $A_1,\dots,A_{n+2}$ satisfy 
\begin{equation}
{\rm eigen}\left({A}_k\right)=(0,{\theta_k})
\quad\hbox{and}\quad
-\sum_{k=1}^{n+2}{A}_k={A}_\infty:=\left(
\begin{array}{ccc}
\rho_1 & & \\ & \dots & \\ & & \rho_{m} 
\end{array}\right).
\label{new}\end{equation}
Moreover $V({\bf t})$ satisfies (\ref{15}) if and only if $A_1,\dots,A_{n+2}$ 
satisfy (\ref{schl}).
\end{thm}

\noindent{\bf Proof.} The proof of this theorem is very similar to the 
proof of Lemmata \ref{lmequiv} and \ref{lmequiv1}. The only difference is that 
instead of dealing with the first $2\times 2$ block with deal with the first 
$m\times m$ block. For example $\Pi$ will be replaced by a diagonal matrix 
with the first $m$ diagonal elements equal to $1$ and the remaining ones equal 
to $0$. {\hfill $\bigtriangleup$}
\vskip 0.2 cm

If in the above theorem we take $m=n+1$, the matrix $V$ has only one null
eigenvalue and the gauge transformation
$$
\tilde Y(z)=z^{-\rho} Y(z),
$$
where $\rho$ is one of the eigenvalues of $V$, is compatible 
with the condition that $V$ has one null eigenvalue. Therefore the above gauge
transformation induces a canonical trasformation of the Schlesinger systems
specified by (\ref{new}) for $m=n+1$.

\begin{rmk}
It is worth observing that our birational canonical transformation $\tilde w$ 
is precisely the one linking the special case PVI$\mu$ of the Painlev\'e
sixth equation studied in \cite{DM,M} and the one studied in \cite{Hit, Hit3}.
The case of PVI$\mu$ is specified by the following choice of the parameters
$$
\alpha={(2\mu-1)^2\over2},\quad\beta=0,\quad\gamma=0,\quad\delta={1\over2},
$$
that is $\theta_1=\theta_2=\theta_3=0 \quad\theta_\infty=2\mu$. Hitchin's case,
say PVI$k$, is specified by 
$$
\alpha={(\sqrt{2 k}-1)^2\over2},\quad\beta=-k,
\quad\gamma=k,\quad\delta={1-2k\over2},
$$ 
that is $\theta_1,\theta_2,\theta_3,\theta_\infty=\pm\sqrt{2 k}$.
Therefore the transformation $\tilde w$ permits to map the classification 
results of \cite{DM,M} to Hitchin's case PVI$k$.
\end{rmk}


\renewcommand{\theequation}{A-\arabic{equation}}
  \setcounter{equation}{0}  
  \section*{APPENDIX}  

We describe the monodromy data of the system (\ref{irreg}). A general 
description of monodromy data of linear systems of ODE can be
found in \cite{MJ1,Mj2,MJU}. The treatment of systems of the form
(\ref{irreg}) can be found in \cite{Dub7} where the case of antisymmetric 
$V$ is dealt with. Here we essentially adapt (omitting all proofs) the 
description of \cite{Dub7} to any $V$ satisfying our conditions (\ref{cV}). 

\vskip 0.4 cm
\noindent{\bf A.1. Local theory at zero}
\vskip 0.4 cm

The system (\ref{irreg}) has a simple pole at $z=0$. We assume that a 
branch-cut between zero and infinity has been chosen along a fixed
line $l$ and a branch of $\log z$ has been selected. 

\begin{prop}
There exists a gauge transformation
$Y={\cal G}(z)\tilde Y$ where ${\mathcal G}(z)=
\sum_{k=0}^\infty{\cal G}_k z^k$, 
convergent near $0$, with principal term ${\cal G}_0:=G$ defined by 
$V={G}\rho {G}^{-1}$, that maps the system (\ref{irreg}) 
into the {\it Birkhoff Normal form:}
\begin{equation}
\ddz\tilde Y=\left({\rho-\ID\over z} + 
\sum_{k\geq1}{\cal R}_k z^{k-1}\right)\tilde Y
\label{N6.2}
\end{equation}
where ${\cal R}=\sum_{k\geq 1} {\cal R}_k$, with
$$
{\cal R}_{k_{ij}}\neq 0,\qquad\iff\qquad \rho_i-\rho_j=k.
$$
As a consequence there exists a fundamental matrix of the system 
(\ref{irreg}) of the form
\begin{equation}
Y_0(z)={\cal G}(z)z^{\rho-\ID} z^{{\cal R}},\quad\hbox{as}\quad 
 z\rightarrow 0.
\label{N6.11}
\end{equation}
\end{prop}

The monodromy ${\cal M}_0$ of the system (\ref{irreg}) with respect to 
the normalized fundamental matrix (\ref{N6.11}) generated by a simple 
closed loop around the origin is 
$$
{\cal M}_0=\exp(2\pi i\rho)\exp(2\pi i{\cal R}).
$$

\vskip 0.4 cm
\noindent{\bf A.2. Local theory at infinity}
\vskip 0.4 cm

\begin{prop} For the system (\ref{irreg}) there exists a unique formal
power series
\begin{equation}
P(z)=\sum_0^\infty P_k z^{-k}
\end{equation}
with $P_0=\ID$, such that the formal gauge transformation
$Y=P(z)\tilde Y$ changes the system (\ref{irreg}) into
the system in normal form:
\begin{equation}
\ddz\tilde Y=\left({\bf T}+{{\bf\Theta}\over z}\right)\tilde Y,\label{S1.5}
\end{equation}
where ${\bf\Theta}$ is a diagonal matrix of entries $-\theta_1-1,\dots,
-\theta_{n+2}-1$.
As a consequence, there is a unique formal fundamental solution 
$Y^\infty_f$ to the system (\ref{irreg}) at 
$z=\infty$\footnote{We fix the branch of the logarithm required in
the definition of $z^\Theta$ as in the previous section.}
\begin{equation}
Y^\infty_f= P(z) z^{\bf\Theta} e^{z {\bf T}}.
\label{S2}\end{equation}
\label{S3}\end{prop}

The above result establishes only the existence of {\it formal
solutions.}\/ Regarding the true solutions, we need to set up 
some more machinery.

\begin{df}
The half-line  
\begin{equation}
R_{ij}=\{z| \RE[z(t_i-t_j)]=0,\,\IM[z(t_i-t_j)]<0\}
\end{equation}
oriented from zero to infinity is called {\it Stokes-ray}.
\label{S3.5}
\end{df}

\begin{df}
An oriented line $l$ in the complex plane is called {\it admissible
 with respect to the points}\/ $(t_1,\dots,t_{n+2})$ if it is such that 
all the Stokes rays $R_{ij}$ with $i<j$ lie on the left of $l$. Vice versa, 
fixed any oriented line $l$ in the complex plane, 
the set of points $(t_1,\dots,t_{n+2})$ is called {\it admissible with
respect to the line $l$,}\/ if all the 
Stokes rays $R_{ij}$ with $i<j$ lie on the left of $l$.
\label{SS3.5}
\end{df}

\begin{df}
(Valid for the particular case we are working with). Let
$\Sigma$ be some sector near $\infty$ and $f(z)$ analytic for $z\in
\Sigma\cap\{|z|>N\}$, $N\in\reali$.
We say that $\sum_{k=0}^n {a_k\over z^k}$ is the {\it asymptotic series
expansion}\/ of $f(z)$ in $\Sigma\cap\{|z|>N\}$,
$$
f(z)\sim\sum_{k=0}^\infty {a_k\over z^k},
$$
if for all $m\in\interi_+$, and for $z\to\infty$ inside $\Sigma$
$$
\lim_{z\to\infty}z^m\left(f(z)-\sum_{k=0}^m {a_k\over z^k}\right)=0.
$$
\label{def5}\end{df}

\begin{lm} Fix an admissible oriented line $l$ and consider the system 
(\ref{irreg}) and its unique formal fundamental solution (\ref{S2}). 
Then there exists $\eps>0$ 
small enough, two sectors $\Pi_L$ and $\Pi_R$ defined as
\begin{equation}
\begin{array}{cc}
&\Pi_R=\{z: \arg(l)-\pi-\eps<\arg(z)<\arg(l)+\eps\}\\
&\Pi_L=\{z: \arg(l)-\eps<\arg(z)<\arg(l)+\pi+\eps\}\\
\end{array}
\end{equation}
and two fundamental solutions $Y_L(z)$ in $\Pi_L$ and
$Y_R(z)$ in $\Pi_R$ such that
\begin{equation}
Y_{L,R}\sim\left(\ID+{\cal O}\left({1\over z}\right)\right)
z^{\bf\Theta} e^{z {\bf T}},\quad\hbox{as }z\to\infty,\quad z\in \Pi_{L,R}.
\label{2.6}
\end{equation}
The fundamental solutions $Y_{L,R}(z)$ are uniquely determined by 
(\ref{2.6}). 
\label{S4}\end{lm}

\noindent The proof can be found in \cite{BJL}.
\vskip 0.1 cm

We are now going to define one of the main objects in the monodromy 
theory of our system at infinity: the Stokes matrices.
In both of the narrow sectors 
$$
\begin{array}{cl}
\Pi_+:=&\{z|\, \phi-\varepsilon< \arg z<\phi+\varepsilon\}\\
\Pi_-:=&\{z|\,\phi-\pi-\varepsilon< \arg z<\phi-\pi+\varepsilon\}\\
\end{array}
$$
obtained by the intersection of $\Pi_l$ and $\Pi_R$, 
we have two fundamental matrices. They must be related by multiplication 
by a constant invertible matrix
$$
Y_L (z) =Y_R (z) S_+, ~~z\in \Pi_+.
$$
$$
Y_L (z) =Y_R (z) S_-, ~~z\in \Pi_-.
$$

\begin{df}
The matrices $S_+$, $S_-$ are called {\it Stokes
matrices of the system (\ref{irreg})}\/ with respect to the admissible
line $\ell$.
\label{def8}\end{df}

\vskip 0.4 cm
\noindent{\bf A.3. Monodromy data}
\vskip 0.4 cm

Resuming, for our system (\ref{irreg})
$$
\ddz Y=\left({\bf T}+{V-\ID\over z} \right) Y,
$$
where ${\bf T}$ is a diagonal matrix with pairwise distinct entries 
$t_1,\dots,t_{n+2}$ and $V$ is diagonalizable with $n$ null eigenvalues, 
we have built three 
distinguished bases in the space  of solutions, i.e. $Y_0(z)$ near
$0$, and $Y_{L,R}(z)$ near $\infty$ depending on the choice of
the admissible line $\ell$. 

To complete the list of the monodromy data we define the {\it central 
connection matrix}\/ between $0$ and $\infty$
$$
Y_0(z) =Y_{R,L}(z)C_{R,L}, ~~z\in \Pi_{R,L}.
$$
Recall that to define $Y_0$ we had to fix the branch cut of
$\log(z)$. We can choose it along the negative part $l_-$ of the 
same line $l$. 

We can reduce the list of the monodromy data $(\rho,R,S_+,S_-,C_R,C_L)$,
by noticing the following {\it cyclic relation}
$$
C_R^{-1}S_-^T S_+^{-1}C_R=C_L^{-1}S_+^T S_-^{-1}C_LM_0 =
\exp(2\pi i\rho)\exp(2\pi{\cal R}).
$$
This expresses a simple topological fact: on the punctured plane 
$\complessi\setminus\{0\}$ a loop around infinity is homotopic to 
a loop around the origin. For a similar reason $C_L=S_+^{-1}C_R$. 

\begin{df}
The {\it Monodromy data}\/ of the system (\ref{irreg}) are a collection of 
constant matrices
$$
\left(\rho,R,S_+,C_R\right)
$$
\label{def9}\end{df}

\begin{thm}
Two systems of the same form (\ref{irreg}) coincide if and only of they have 
the same monodromy data.
\label{th4}\end{thm}

The proof can be found in \cite{Dub7}.

\bibliography{bibliografia.bib}
\bibliographystyle{plain}

\end{document}